\documentclass[conference]{IEEEtran}

\usepackage[pdftex]{graphicx}
\graphicspath{{fig/}}
\DeclareGraphicsExtensions{.pdf,.jpeg,.png}

\usepackage{url}
\usepackage{colortbl} % For cell colors
\usepackage{xcolor}   % For color gradients

\usepackage{enumitem}

\usepackage{amsmath}
\usepackage{amssymb}

\usepackage{subcaption}
\usepackage{tikz}

\usepackage[utf8]{inputenc}
\usepackage{multirow, booktabs}

\usepackage{siunitx}

\usepackage[
 backend=biber
,style=ieee
,minbibnames=1
,maxbibnames=2
,maxcitenames=1
,mincitenames=1
,eprint=false
,doi=true
,isbn=false
,url=true
]{biblatex}
\bibliography{ref} 
\AtBeginBibliography{\footnotesize}

\DeclareSourcemap{
  \maps{
    \map{
      \step[fieldset=month, null]
      \step[fieldset=address, null]
      \step[fieldset=location, null]
      \step[fieldset=publisher, null]
      \step[fieldset=url, null]
      \step[fieldset=isbn, null]
    }
  }
}

\def\BibTeX{{\rm B\kern-.05em{\sc i\kern-.025em b}\kern-.08em
    T\kern-.1667em\lower.7ex\hbox{E}\kern-.125emX}}

\usepackage{cleveref}
\Crefname{figure}{Fig.}{Figs.}

\usepackage{flushend}

\begin{document}
%
% paper title
\title{AutoFlow: An Autoencoder-based Approach for IP Flow Record Compression with Minimal Impact on Traffic Classification}

\author{
    \IEEEauthorblockN{Adrian Pekar}
    \IEEEauthorblockA{
        Department of Networked Systems and Services, Faculty of Electrical Engineering and Informatics,\\ 
        Budapest University of Technology and Economics, M\H{u}egyetem rkp. 3., H-1111 Budapest, Hungary.\\
        HUN-REN-BME Information Systems Research Group, Magyar Tud\'{o}sok krt. 2, 1117 Budapest, Hungary.\\
        CUJO LLC, Budapest, Hungary.\\
        Email: apekar@hit.bme.hu
    }
}

\newpage

% make the title area
\maketitle

\begin{abstract}
Network monitoring generates massive volumes of IP flow records, posing significant challenges for storage and analysis. This paper presents a novel deep learning-based approach to compressing these records using autoencoders, enabling direct analysis of compressed data without requiring decompression. Unlike traditional compression methods, our approach reduces data volume while retaining the utility of compressed data for downstream analysis tasks, including distinguishing modern application protocols and encrypted traffic from popular services.
Through extensive experiments on a real-world network traffic dataset, we demonstrate that our autoencoder-based compression achieves a $1.313\times$ reduction in data size while maintaining $99.27\%$ accuracy in a multi-class traffic classification task, compared to $99.77\%$ accuracy with uncompressed data. This marginal decrease in performance is offset by substantial gains in storage and processing efficiency. The implications of this work extend to more efficient network monitoring and scalable, real-time network management solutions.
\end{abstract}

\begin{IEEEkeywords}Network Traffic Analysis, IP Flow Compression, Deep Learning, Autoencoders, Traffic Classification
\end{IEEEkeywords}

\begin{tikzpicture}[remember picture,overlay]
\node[anchor=north, align=center, text=red, font=\small, yshift=-.6cm] at (current page.north) {This version of the paper was accepted for presentation at the 2025 IEEE/IFIP Network Operations and Management Symposium (NOMS 2025)};
\end{tikzpicture}

\IEEEpeerreviewmaketitle

\section{Introduction}

Network traffic analysis and classification play crucial roles in modern network management, security, and quality of service provisioning. As network infrastructures grow in complexity, the volume of data generated by monitoring systems has increased exponentially \cite{cisco2020annual}, creating significant challenges for storage and analysis of IP flow records \cite{Hofstede2014,Sadre2014}. These challenges are particularly acute for real-time tasks such as anomaly detection and performance monitoring \cite{Boutaba2018}.

To address these challenges, there is a pressing need for efficient data compression techniques that can significantly reduce the storage footprint of IP flow records while preserving their utility for downstream analysis tasks. Traditional compression methods often fail to capture the inherent structure and relationships within network traffic data, potentially leading to loss of critical information for analysis \cite{Tongaonkar2015}.

We propose an autoencoder-based approach to IP flow record compression that leverages neural networks to learn compact, low-dimensional representations while preserving essential characteristics for accurate traffic classification. By doing so, we aim to achieve a reduction in data volume while maintaining high performance in downstream analysis tasks.

Our experimental results demonstrate the effectiveness of this approach, achieving a compression ratio of $1.313\times$ while maintaining $99.27\%$ accuracy in multi-class traffic classification, compared to $99.77\%$ with uncompressed data. The method shows particular promise in distinguishing between various modern application protocols, including encrypted traffic.

The implications of this work are far-reaching, potentially enabling more efficient storage and processing of network monitoring data, facilitating real-time analysis, and paving the way for more scalable network management solutions. Furthermore, the compressed representations learned by our model may offer new insights into the underlying structure of network traffic, potentially leading to improved analysis techniques.

The remainder of this paper is organized as follows: 
\Cref{sec:related_work} discusses related work relevant in the context of this work. 
\Cref{sec:methodology} details our proposed autoencoder-based compression method.  
\Cref{sec:results} presents and analyzes our results. 
\Cref{sec:discussion} discusses the implications and future research directions.
\Cref{sec:conclusion} concludes the paper. 

\section{Related Work} 
\label{sec:related_work}

Data compression in network traffic has been extensively studied to improve bandwidth utilization, reduce storage requirements, and enhance processing efficiency. 

Header compression techniques aim to reduce overhead for efficient bandwidth usage. For example, \citeauthor{Farouq2020}~\cite{Farouq2020} proposed methods to compress headers in real-time video streaming over UDP/IP and HTTP/TCP flows, achieving header size reductions of up to $90\%$ for RTP/UDP/IP. \citeauthor{Westphal2003-1}~\cite{Westphal2003-1,Westphal2003-2} developed schemes exploiting similarities in consecutive packet headers to enhance bandwidth utilization in wireless networks by combining time and space compression algorithms.

\citeauthor{Sonai2024WSN}~\cite{Sonai2024WSN} introduced a statistical compression method for wireless sensor networks, achieving a $91.23\%$ memory reduction and $79.65\%$ energy savings through normalization and encoding/decoding algorithms. 
\citeauthor{Sonai2024CTLA}~\cite{Sonai2024CTLA} also developed the Compressed Table Look-up Algorithm for OpenFlow switches, which combines Huffman encoding with a modified trie structure, hashing, and recursive binary search to optimize IPv4 and IPv6 lookups, achieving a $37\%$ and $61\%$ space reduction, respectively, and improving lookup time complexity~\cite{Sonai2024CTLA}. 

For packet-level compression, \citeauthor{Chen2008}~\cite{Chen2008} developed IPzip exploiting packet correlations, while \citeauthor{Huang2014}~\cite{Huang2014} achieved $70\%$ reduction using memory-assisted clustering algorithms.

While these studies offer valuable insights, they focus on packet-level or device-specific compression rather than IP flow records---aggregated summaries essential for network monitoring. Our work addresses this gap through an autoencoder-based method that learns compact representations while preserving classification utility, providing a foundation for scalable network monitoring solutions.

\section{Methodology}
\label{sec:methodology}

This section details our autoencoder-based method for IP flow record compression and its evaluation through traffic classification tasks.

\subsection{Dataset Description}
\label{sec:dataset}

Our study uses a dataset of \num{3163140} network flows collected from a university dormitory network, with implementation details and data access documented in~\cite{Pekar2024}. The dataset contains \num{91} flow features~\cite{nfs-core-feat}, spanning volume and count metrics, statistical summaries (minimum, mean, maximum, and standard deviation) of packet lengths and inter-arrival times, and application characteristics. It covers diverse applications including web traffic, streaming services, social media, messaging apps, file transfers, and remote desktop protocols.

\subsection{Feature Selection for Compression}
\label{sec:feature_selection}

Feature selection for compression requires careful consideration of reconstruction error impact. For instance, reconstructing a flow size of $225$ MB as $222$ MB may be acceptable in some contexts, whereas reconstructing a timestamp incorrectly can lead to major inaccuracies in event sequencing and analysis. Critical features that must be preserved exactly include timestamps (for event sequencing), IP addresses (endpoint identification), port numbers (service identification), and protocol information (traffic interpretation). For compression, we selected $21$ features across four categories where minor reconstruction errors are tolerable: bidirectional metrics (duration, packet/byte counts), source-to-destination metrics, destination-to-source metrics, and statistical packet size metrics (min, mean, std, max) for both directional and bidirectional flows.

\subsection{Data Preprocessing}
\label{subsec:preprocessing}

Ou data preprocessing involves two key steps:

\begin{enumerate}

\item \textbf{Outlier handling} through a $99.9$th percentile clipping strategy for each feature:
\begin{equation}
    f_i' = \min(f_i, p_{99.9}(f_i)).
\end{equation}
This approach affects only $0.1\%$ of the data points while stabilizing model training~\cite{Hodge2004}. We specifically chose this high threshold to preserve the heavy-tailed characteristics common in network traffic, where extreme events often represent important phenomena like traffic bursts or potential anomalies. 

\item \textbf{Robust scaling} using the inter-quartile range:
\begin{equation}
    f_i'' = \frac{f_i' - \text{median}(f_i)}{\text{IQR}(f_i)}.
\end{equation}
This approach is less sensitive to outliers compared to standard scaling methods, making it particularly suitable for network traffic data which often contains anomalies~\cite{rousseeuw2011robust}.

\end{enumerate}

\subsection{Autoencoder Architecture}
\label{subsec:autoencoder}

Our autoencoder architecture consists of an encoder $E$ and a decoder $D$:
\begin{equation}
E: \mathbb{R}^n \rightarrow \mathbb{R}^m, \quad D: \mathbb{R}^m \rightarrow \mathbb{R}^n,
\end{equation}
where $n=21$ input features are compressed to $m=16$ dimensions. The bottleneck layer of 16 neurons, determined through empirical experiments, provides optimal balance between compression and reconstruction quality.

Our architecture comprises:
\begin{itemize}
    \item \textbf{Encoder:} Three fully connected layers ($128$, $64$, $16$ neurons)
    \item \textbf{Decoder:} Three fully connected layers ($64$, $128$, $21$ neurons)
\end{itemize}

The encoder and decoder are formulated as:
\begin{equation}
\begin{split}
E(\mathbf{x}) &= \sigma(W_3\sigma(W_2\sigma(W_1\mathbf{x} + b_1) + b_2) + b_3), \\
D(\mathbf{z}) &= W_6\sigma(W_5\sigma(W_4\mathbf{z} + b_4) + b_5) + b_6,
\end{split}
\end{equation}
where $W_i$ are weight matrices, $b_i$ are bias vectors, and $\sigma$ is the activation function. 

Through preliminary experiments comparing various activation functions, we selected LeakyReLU as the activation function, with a negative slope of 0.2:
\begin{equation}
\sigma(x) = \max(0.2x, x).
\end{equation}

LeakyReLU helps prevent the "dying ReLU" problem, allowing for more nuanced feature learning \cite{maas2013rectifier}.

\subsection{Training Process}
\label{subsec:training}

Our training process incorporates several techniques to ensure robust and reproducible results:

\begin{enumerate}
    \item \textbf{Loss Function:} We use Huber loss, defined as:
    
    \begin{equation}
        L_\delta(y, f(x)) = \begin{cases}
            \frac{1}{2}(y - f(x))^2, & \text{for } |y - f(x)| \leq \delta, \\
            \delta(|y - f(x)| - \frac{1}{2}\delta), & \text{otherwise},
        \end{cases}
    \end{equation}
    
    where $\delta = 1$ in our implementation. Huber loss combines the best properties of Mean Squared Error (MSE) for small errors and Mean Absolute Error (MAE) for large errors, making it robust to outliers \cite{Huber1964}.
    
    \item \textbf{Optimizer:} Adam optimizer with a learning rate of $0.001$ and weight decay of \num{1e-5} for regularization. Adam is chosen for its ability to adapt the learning rate for each parameter, which is particularly useful for training deep neural networks \cite{Kingma2014}.
    
    \item \textbf{Learning Rate Scheduling:} We implement a ReduceLROnPlateau scheduler, which reduces the learning rate by a factor of $0.5$ when the validation loss plateaus for $5$ epochs. This adaptive approach helps in fine-tuning the model and avoiding local optima \cite{Smith2017}.
    
    \item \textbf{Gradient Clipping:} We apply gradient clipping with a maximum norm of $1.0$ to prevent exploding gradients, a common issue in training deep networks \cite{Pascanu2013}.

    \item \textbf{Training:} The autoencoder is trained to minimize the reconstruction loss:

    \begin{equation}
    \min_{E,D} \mathbb{E}_{\mathbf{x} \sim P_{data}}[L(\mathbf{x}, D(E(\mathbf{x})))],
    \end{equation}
    where $P_{data}$ is the empirical distribution of the preprocessed IP flow records.
    
    \item \textbf{Training Duration:} The model is trained for a maximum of $200$ epochs with early stopping based on validation loss to prevent overfitting.

    \item \textbf{Reproducibility:} For reproducibility, all implementation details and random seeds are provided in our digital artifact~\cite{github}.

\end{enumerate}

\subsection{Compression Evaluation}
\label{subsec:compression_evaluation}

We evaluate our method using complementary metrics that capture different aspects of compression performance and carefully account for details that affect measurement accuracy:

\begin{itemize}
    \item \textbf{Compression Ratio:} For compression efficiency evaluation, we consider several important perspectives. 
        \begin{itemize}
            \item \textit{Naive ratio:} measures raw dimensionality reduction as:
            \begin{equation}
                \text{CR}_{\text{naive}} = \frac{\text{Size}(\mathbf{X}_{\text{train}})}{\text{Size}(\mathbf{Z}_{\text{train}})},
            \end{equation}
            where $\mathbf{X}_{\text{train}}$ represents the training data and $\mathbf{Z}_{\text{train}}$ represents its latent space encoding.
            
            \item \textit{Pure ratio:} The naive metric combines both dimensionality reduction ($21$ to $16$ features) and an unintended precision change from scikit-learn's \texttt{float64} to PyTorch's \texttt{float32}. To isolate the true dimensionality reduction effect, we compute a pure compression ratio using consistent \texttt{float32} precision:
            \begin{equation}
                \text{CR}_{\text{pure}} = \frac{\text{Size}(\mathbf{X}_{\text{train}}^{\text{f32}})}{\text{Size}(\mathbf{Z}_{\text{train}}^{\text{f32}})}.
            \end{equation}

            \item \textit{Practical ratio:} For real-world deployment considerations, we include the autoencoder model's storage overhead:
            \begin{equation}
                \text{CR}_{\text{practical}} = \frac{\text{Size}(\mathbf{X}_{\text{train}}^{\text{f32}})}{\text{Size}(\mathbf{Z}_{\text{train}}^{\text{f32}}) + \text{Size}(\text{Model})}.
            \end{equation}

            \item \textit{Benchmark ratio}: To contextualize our autoencoder's performance, we benchmark against traditional compression methods using:
            \begin{equation}
                \text{CR}_{\text{method}} = \frac{\text{Size}(\mathbf{X}_{\text{train}}^{\text{f32}})}{\text{Size}(\text{Compressed}_{\text{method}})},
            \end{equation}
            where method $\in \{\text{ZIP}, \text{LZMA}\}$, though these methods require decompression before analysis.
    \end{itemize}
    
    \item \textbf{Reconstruction Error:} We measure the fidelity of the reconstructed data using:
    \begin{itemize}
        \item Root Mean Squared Error (RMSE):
        \begin{equation}
            \text{RMSE} = \sqrt{\frac{1}{n} \sum_{i=1}^n (x_i - \hat{x_i})^2}.
        \end{equation}
        \item Mean Absolute Percentage Error (MAPE):
        \begin{equation}
            \text{MAPE} = \frac{100\%}{n} \sum_{i=1}^n \left|\frac{x_i - \hat{x_i}}{x_i}\right|,
        \end{equation}
        where $x_i$ is the original value and $\hat{x_i}$ is the reconstructed value.
    \end{itemize}

    \item \textbf{Feature-wise Median Percentage Error:} For each feature $j$, we compute the median of the absolute percentage differences between original and reconstructed values:
    \begin{equation}
        \text{MdPE}_j = \text{median}\left(\left|\frac{X_j - \hat{X}_j}{X_j}\right| \times 100\right),
    \end{equation}
    where $X_j$ represents the original values for feature $j$ and $\hat{X}_j$ represents the reconstructed values for feature $j$. 
    % Note that a small constant ($\epsilon = 10^{-8}$) is added to the denominator to prevent division by zero.
    
    \item \textbf{Distribution Preservation:} We evaluate distribution similarity using Kullback-Leibler (KL) Divergence with continuous density estimation:
    \begin{equation}
        D_{KL}(P||Q) = \int_{-\infty}^{\infty} p(x) \log\left(\frac{p(x)}{q(x)}\right) dx,
    \end{equation}
    where $p$ and $q$ are probability density functions of original and reconstructed features estimated using Gaussian Kernel Density Estimation, with the bandwidth selected using Scott's rule. The integral is evaluated numerically using Simpson's rule, with adaptive numerical stability handling.
    % This metric is particularly important for heavy-tailed network traffic features, where preserving distribution characteristics is crucial for downstream analysis.

    \item \textbf{Feature Correlation Preservation:} We compute the difference in correlation matrices between original and reconstructed features to assess the preservation of feature relationships:
    \begin{equation}
        \Delta\text{Corr} = \text{Corr}(X) - \text{Corr}(\hat{X}),
    \end{equation}
    where $X$ is the matrix of original features and $\hat{X}$ is the matrix of reconstructed features.
\end{itemize}

\subsection{Traffic Classification}
\label{subsec:classification}

To assess the practical utility of our compression method, we evaluate its impact on traffic classification performance. Our evaluation framework consists of two parallel classification pipelines---one using original features and another using compressed representations---allowing direct comparison of classification efficacy.

For compressed data classification, we extract 16-dimensional latent representations using the trained encoder, while original data classification uses the full $21$ features. Both pipelines employ a Random Forest classifier with $100$ estimators, chosen for its robustness to non-linear relationships and overfitting resistance~\cite{Breiman2001}. We maintain identical training conditions by using stratified $80$-$20$ train-test splits to ensure balanced class distributions.

Our evaluation employs complementary metrics to thoroughly assess classification performance:

\begin{itemize}
    \item \textbf{Overall Performance:} We measure accuracy as the primary metric, supplemented by macro-averaged and weighted-averaged F1-scores to account for potential class imbalances.
        
    \item \textbf{Comparative Analysis:} We compare the classification performance between the original and compressed features to assess the impact of our compression method on the downstream task.
    
    \item \textbf{Misclassification Analysis:} We perform a detailed analysis of misclassifications to understand the strengths and limitations of our compressed representations compared to the original features.
    
\end{itemize}

This comprehensive evaluation framework, with implementation details available in our digital artifact~\cite{github}, enables assessment of both compression effectiveness and its practical impact on traffic classification capability.

\section{Results}
\label{sec:results}

This section presents our experimental results, evaluating both the compression performance and its impact on the downstream task of traffic classification. 

\subsection{Autoencoder Performance}
\label{subsec:autoencoder_performance}

Table~\ref{tab:com_perf} presents our compression performance metrics, revealing important insights about the practical efficiency of our approach. Notably, while our naive compression ratio suggests a $2.625\times$ reduction, accounting for implementation details provides a more nuanced understanding of actual performance.

\begin{table}[h!]
    \centering
    \caption{Compression Performance}
    \label{tab:com_perf}
    \begin{tabular}{@{}lr@{}}
        \toprule
        \textbf{Metric}            & \textbf{Value}        
        \\ \midrule
        $\text{CR}_{\text{naive}}$ & $2.625$ \\
        $\text{CR}_{\text{pure}}$ & $1.313$ \\
        $\text{CR}_{\text{practical}}$ & $1.312$ \\
        \midrule
        $\text{CR}_{\text{ZIP}}$ & $1.498$ \\
        $\text{CR}_{\text{LZMA}}$ & $2.427$ \\
        \midrule
        RMSE                       & \num{37130}           \\ 
        MAPE                       & \num{3071189490}$\%$    \\
        \bottomrule
    \end{tabular}
\end{table}

The pure compression ratio of $1.313\times$ reflects our architectural choice of reducing $21$ features to $16$ dimensions, while the practical ratio of $1.312\times$ accounts for model storage overhead. 
The pure compression ratio indicates that our model successfully reduced the data volume to approximately $76\%$ of its original size, showcasing its effectiveness in data compression.
Interestingly, our method achieves comparable compression to ZIP ($1.498\times$) while enabling direct analysis of compressed data, though LZMA achieves higher compression ($2.427\times$) at the cost of requiring decompression before use.

The reconstruction metrics require careful interpretation within the context of network traffic characteristics. The high RMSE (\num{37130}) reflects the scale of our features. For instance, a flow with $1$ billion bytes reconstructed as 0.99 billion bytes would be highly accurate in relative terms but contribute significantly to RMSE. Similarly, the large MAPE value primarily results from near-zero values in the original data, where small absolute differences translate to large percentage errors. 

\subsection{Feature-specific Reconstruction}
\label{subsec:feature_reconstruction}

\newcommand{\errorcolor}[1]{%
    \ifdim#1pt>2pt
        \cellcolor{red!50}%
    \else\ifdim#1pt>1.5pt
        \cellcolor{red!40}%
    \else\ifdim#1pt>1.0pt
        \cellcolor{yellow!50}%
    \else\ifdim#1pt>0.5pt
        \cellcolor{yellow!20}%
    \else
        \cellcolor{green!10}%
    \fi\fi\fi\fi
    #1%
}

\newcommand{\klcolor}[1]{%
    \ifdim#1pt>0.005pt
        \cellcolor{blue!50}%
    \else\ifdim#1pt>0.001pt
        \cellcolor{blue!30}%
    \else\ifdim#1pt>0.0001pt
        \cellcolor{blue!20}%
    \else
        \cellcolor{blue!10}%
    \fi\fi\fi
    #1%
}

\begin{table}[!t]
\centering
\caption{Feature-specific Reconstruction Performance}
\label{tab:feature_reconstruction}
\begin{tabular}{lcc}
\toprule
\textbf{Feature} & \textbf{Median Percentage Error} & \textbf{KL Divergence} \\
\midrule
dst2src\_stddev\_ps & \errorcolor{2.344317} & \klcolor{0.001889} \\
src2dst\_stddev\_ps & \errorcolor{2.218252} & \klcolor{0.001478} \\
dst2src\_duration\_ms & \errorcolor{1.775878} & \klcolor{0.000042} \\
bidirectional\_stddev\_ps & \errorcolor{1.611596} & \klcolor{0.000399} \\
src2dst\_duration\_ms & \errorcolor{1.537608} & \klcolor{0.000035} \\
dst2src\_mean\_ps & \errorcolor{1.326498} & \klcolor{0.000314} \\
src2dst\_max\_ps & \errorcolor{1.219881} & \klcolor{0.007644} \\
bidirectional\_duration\_ms & \errorcolor{1.158148} & \klcolor{0.000040} \\
bidirectional\_mean\_ps & \errorcolor{1.127692} & \klcolor{0.000279} \\
dst2src\_max\_ps & \errorcolor{1.018367} & \klcolor{0.002445} \\
dst2src\_bytes & \errorcolor{0.895472} & \klcolor{0.000000} \\
bidirectional\_packets & \errorcolor{0.807773} & \klcolor{0.000453} \\
bidirectional\_max\_ps & \errorcolor{0.740957} & \klcolor{0.005745} \\
src2dst\_mean\_ps & \errorcolor{0.620547} & \klcolor{0.000051} \\
dst2src\_packets & \errorcolor{0.609727} & \klcolor{0.000012} \\
src2dst\_bytes & \errorcolor{0.555504} & \klcolor{0.000001} \\
bidirectional\_bytes & \errorcolor{0.540620} & \klcolor{0.000001} \\
src2dst\_packets & \errorcolor{0.505411} & \klcolor{0.000002} \\
src2dst\_min\_ps & \errorcolor{0.040550} & \klcolor{0.000203} \\
bidirectional\_min\_ps & \errorcolor{0.022322} & \klcolor{0.000062} \\
dst2src\_min\_ps & \errorcolor{0.012635} & \klcolor{0.000004} \\
\bottomrule
\end{tabular}
\end{table}

\begin{figure}[!t]
    \centering
    \includegraphics[width=1\columnwidth]{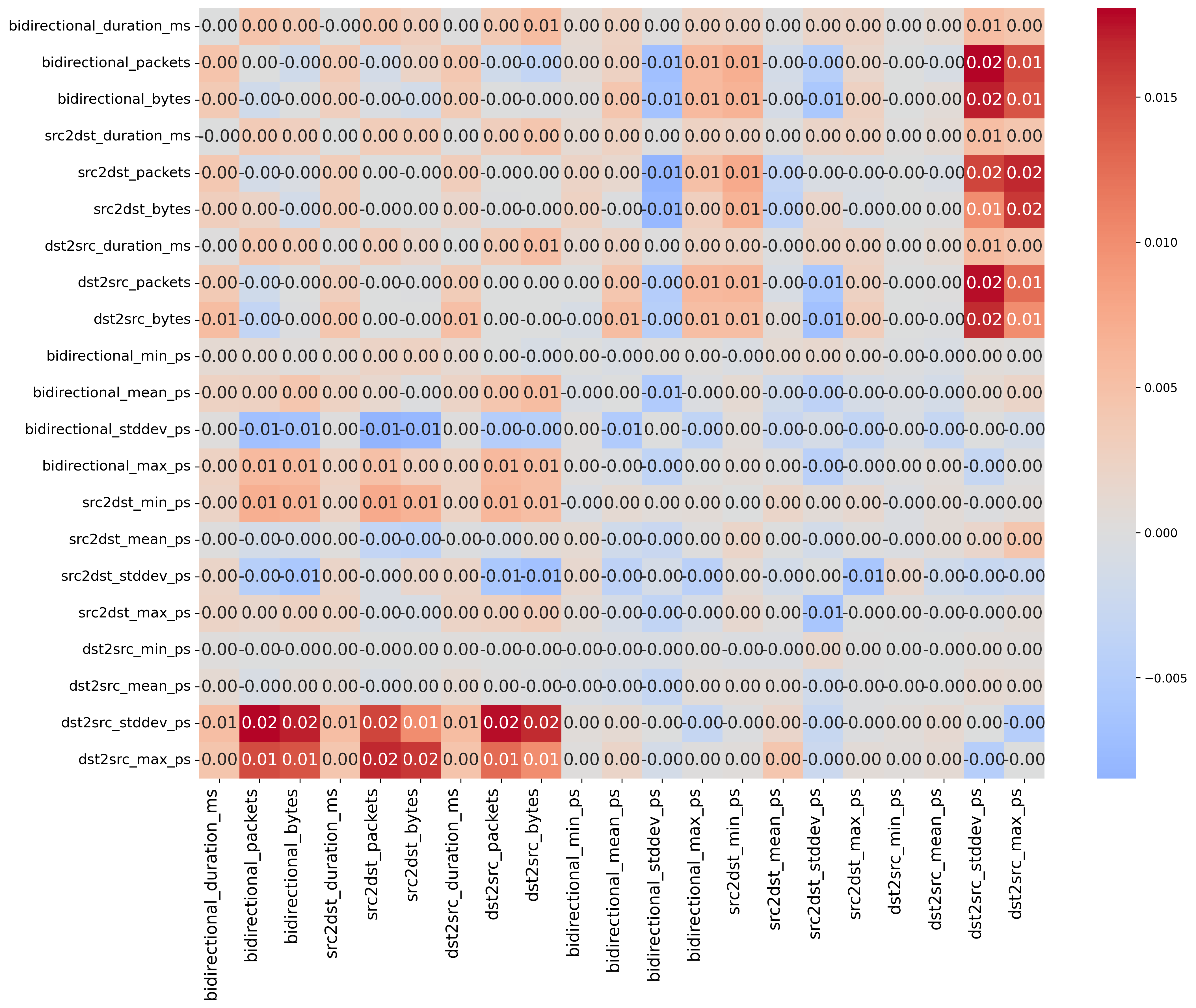}
    \caption{Correlation Difference: Original - Reconstr. Features}
    \label{fig:correlation_diff}
\end{figure}

To understand our model's performance at a granular level, we analyzed reconstruction accuracy using feature-wise median percentage error and KL divergence. \Cref{tab:feature_reconstruction} presents these results sorted by median percentage error.
Our analysis reveals several key patterns in reconstruction performance:

\begin{itemize}
    \item \textbf{Overall Accuracy:} All features maintain median percentage errors below $2.5\%$, with many below $1\%$, indicating strong reconstruction fidelity. The lowest errors ($< 0.05\%$) occur in minimum packet size features, while standard deviation features show higher errors ($\sim2.3\%$), reflecting their inherent variability.
    
    \item \textbf{Distribution Preservation:} KL divergence values range from $0$ to $0.008$, indicating excellent preservation of feature distributions. Byte counts and basic packet metrics show near-perfect preservation (KL $< 0.000002$), while duration-related features maintain moderate divergences (around $0.00004$). \texttt{src2dst\_max\_ps} and \texttt{bidirectional\_max\_ps} show higher divergences ($\sim0.007$), reflecting the greater challenge of modeling the variability in these distributions. However, even the highest KL divergences remain small in absolute terms, underscoring the model's strong overall performance.

    \item \textbf{Feature Correlations:} The correlation difference heatmap in \Cref{fig:correlation_diff} shows mostly small differences ($-0.01$ to $0.01$), with some notable patterns:
    \begin{itemize}
        \item \texttt{dst2src\_stddev\_ps} shows the largest correlation differences (up to $0.02$) with basic flow metrics.
        \item Minimum packet size features maintain extremely stable correlations.
        \item Packet statistics show small but consistent pattern differences.
    \end{itemize}
\end{itemize}
    
These results demonstrate our model's strong overall performance while highlighting specific areas where reconstruction accuracy varies by feature type. The slightly higher errors in standard deviation features suggest potential areas for future architectural optimization, particularly for applications where flow variability metrics are crucial.

\subsection{Traffic Classification Performance}
\label{subsec:classification_performance}

To evaluate practical utility, we compared classification performance between original and compressed features. The compressed representations maintain remarkably high accuracy, as shown in \Cref{tab:classification_performance}.

\begin{table}[!t]
\centering
\caption{Classification Performance Comparison}
\label{tab:classification_performance}
\begin{tabular}{@{}lccc@{}}
\toprule
\multirow{2}{*}{\textbf{Metric}} & \textbf{Original} & \textbf{Compressed} & \multirow{2}{*}{\textbf{Difference}} \\
 & \textbf{Features} & \textbf{Features} & \\
\midrule
Accuracy & $0.997732$ & $0.992677$ & $-0.005055$ \\
Macro Avg F1-score & $0.997826$ & $0.993106$ & $-0.004720$ \\
Weighted Avg F1-score & $0.997731$ & $0.992661$ & $-0.005070$ \\
\bottomrule
\end{tabular}
\end{table}
The classifier using compressed features achieves $99.27\%$ accuracy, compared to $99.77\%$ with original features. This minimal reduction ($0.51$ percentage points) demonstrates strong preservation of discriminative information, supported by similarly small decreases in macro and weighted-average F1-scores.

Our class-specific analysis, detailed in our digital artifact~\cite{github}, reveals that most traffic classes maintain near-perfect classification performance even with compressed features. The largest performance drops occur in TLS.Facebook (F1-score from $0.99$ to $0.97$), while QUIC.Instagram and HTTP show smaller decreases (F1-scores from $1.00$ to $0.99$).

\begin{figure*}[!t]
    \centering
    \begin{subfigure}{.47\textwidth}
        \includegraphics[width=\linewidth]{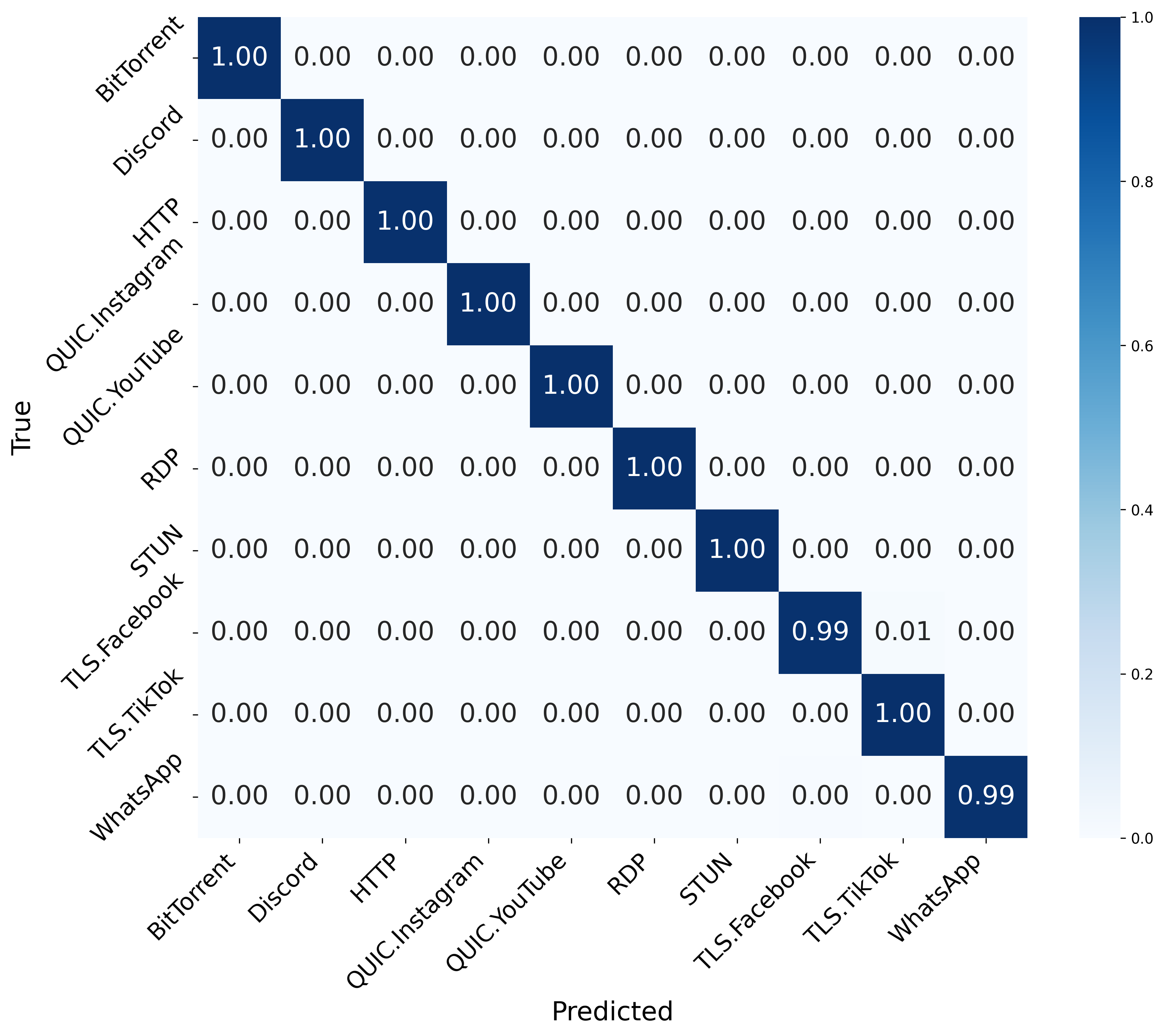}
        \caption{Original Features}
    \end{subfigure}
    \begin{subfigure}{.47\textwidth}
        \includegraphics[width=\linewidth]{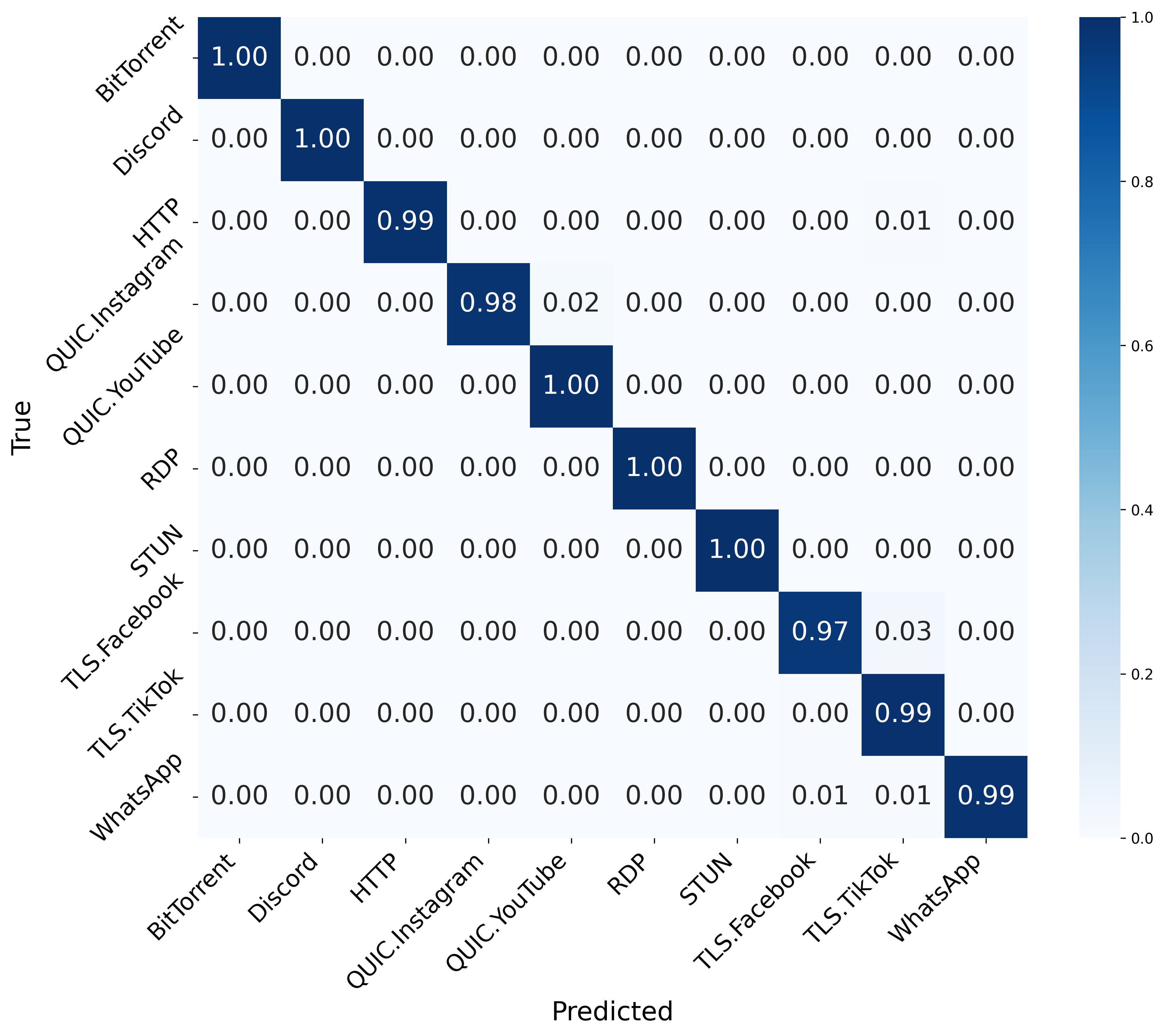}
        \caption{Compressed Features}
    \end{subfigure}
    \caption{Normalized Confusion Matrices for Original and Compressed Features}
    \label{fig:confusion_matrices}
\end{figure*}

The normalized confusion matrices in \Cref{fig:confusion_matrices} reveal subtle changes in classification patterns after compression. With original features, we observe minimal confusion mainly between TLS.Facebook and TLS.TikTok ($0.01$). Compressed features introduce additional minor confusions: QUIC.Instagram with QUIC.YouTube ($0.02$), HTTP with TLS.TikTok ($0.01$), increased confusion between TLS.Facebook and TLS.TikTok ($0.03$), and slight confusion between WhatsApp and TLS services ($0.01$ each).

\begin{table}[!t]
\centering
\caption{Misclassification Analysis}
\label{tab:misclassification}
\begin{tabular}{@{}lcc@{}}
\toprule
\textbf{Metric} & \textbf{Original Features} & \textbf{Compressed Features} \\
\midrule
Total Misclassifications & \num{1435} & \num{4633} \\
\midrule
\multicolumn{3}{@{}l}{\textbf{Top 5 Misclassified Classes:}} \\
\midrule
1 & TLS.Facebook (\num{682}) & TLS.Facebook (\num{2202}) \\
2 & TLS.TikTok (\num{299}) & TLS.TikTok (\num{962}) \\
3 & HTTP (\num{261}) & HTTP (\num{569}) \\
4 & WhatsApp (\num{128}) & QUIC.Instagram (\num{442}) \\
5 & BitTorrent (\num{28}) & WhatsApp (\num{184}) \\
\bottomrule
\end{tabular}
\end{table}

Detailed misclassification summary in \Cref{tab:misclassification} shows compression increases total errors by a factor of $3.2$ (from \num{1435} to \num{4633}). TLS.Facebook and TLS.TikTok remain the most challenging classes to distinguish, with their errors increasing proportionally. QUIC.Instagram emerges as more problematic in the compressed version with $442$ misclassifications, replacing BitTorrent in the top 5 misclassified classes. Despite this increase in errors, the overall accuracy remains high at $99.27\%$, indicating that our compression method effectively preserves discriminative features for most traffic types while introducing minor challenges in distinguishing between similar encrypted protocols.

\section{Discussion}
\label{sec:discussion}

This section analyzes the broader implications of our autoencoder-based compression approach for network traffic analysis.

\subsection{Advantages over Traditional Compression Methods}

While our method achieves lower compression ratios compared to traditional approaches like LZMA ($2.427\times$), it offers distinct advantages for network traffic analysis. Traditional compression methods require complete decompression before any analysis can be performed, whereas our autoencoder enables direct analysis of the compressed representations. This capability becomes particularly valuable in scenarios requiring repeated analysis of the same data.

The overhead characteristics also differ fundamentally. Traditional methods require decompression overhead for each analysis task, with each compressed file containing its own header information. Our approach shifts this overhead to a one-time model training phase, after which the same model can compress multiple datasets. This amortization becomes especially beneficial in large-scale deployments processing numerous flow records.

Another key distinction lies in the nature of compression. Traditional methods offer lossless compression but require complete decompression for any analysis. Our approach makes an intentional trade-off: accepting minor reconstruction errors in carefully selected features while maintaining high fidelity for downstream tasks. With classification accuracy only dropping from $99.77\%$ to $99.27\%$, this trade-off enables a new workflow where analysis can be performed directly on compressed data.

\subsection{Limitations and Future Directions}

Building on our promising results, we identify several limitations and areas for future work:

% Our work establishes a foundation for autoencoder-based network flow compression while highlighting several promising directions for future research:
% Our study reveals several important limitations and areas for future work:

\begin{itemize}
    \item \textbf{Architecture Optimization:} While our initial architecture choice of 16 hidden units proved effective, this was not systematically optimized. A comprehensive exploration of different architectures and hyperparameters could potentially yield better compression-accuracy trade-offs. 
    % This includes investigating the impact of layer sizes, activation functions, and optimization strategies on both compression ratio and classification performance.
    
    \item \textbf{Architecture Exploration:} While we focused on a vanilla autoencoder, our experiments with a denoising variant (available in our digital artifact~\cite{github}) showed the same compression ratio but significantly higher reconstruction errors and lower classification accuracy, suggesting that the additional noise resistance introduces unnecessary information loss for our use case.
    Future work could investigate more advanced architectures. 
    % Particularly, architectures designed for sequence data might better capture the temporal aspects of network flows.

    % \item \textbf{Feature Distribution:} Our results show higher KL divergences for packet size statistics, suggesting room for improvement in preserving complex distributions. Future work could explore specialized loss functions that better preserve these statistical properties while maintaining compression efficiency.
    
    \item \textbf{Encrypted Traffic:} The increased confusion between similar TLS services indicates a need for better feature preservation strategies for encrypted traffic. This becomes increasingly important as encrypted traffic continues to dominate network communications. Potential approaches could include targeted feature selection or specialized architectures for encrypted flow characteristics.

\end{itemize}

\section{Conclusion}
\label{sec:conclusion}

This paper presented a novel approach to IP flow record compression using autoencoders, demonstrating the feasibility of maintaining high classification accuracy while reducing data volume. Our method achieved a practical compression ratio of $1.312\times$ when accounting for implementation overhead, comparable to traditional compression methods ($1.498\times$ for ZIP) while enabling direct analysis of compressed data. The compressed representations maintained a classification accuracy of $99.27\%$, compared to $99.77\%$ with original features, across a diverse set of network applications including encrypted traffic. This minimal performance drop, coupled with the ability to perform analysis without decompression, offers a practical solution for network monitoring scenarios requiring repeated data analysis.

\section*{Acknowledgment}

Supported by the János Bolyai Research Scholarship of the Hungarian Academy of Sciences. 
% This work was supported by Project no. 2024-1.2.6-EUREKA-2024-00009. 
Project no. 2024-1.2.6-EUREKA-2024-00009 has been implemented with the support provided by the Ministry of Culture and Innovation of Hungary from the National Research, Development and Innovation Fund, financed under the 2024-1.2.6-EUREKA funding scheme.

\printbibliography

\end{document}